\documentclass[twocolumn,floats,showpacs,superscriptaddress,pre]{revtex4}
\usepackage{amsmath,amssymb,bm}
\usepackage{graphics}
\usepackage{epsfig}
\usepackage{graphicx}

\begin{document}

\title{Quantum Action Principle in Relativistic Mechanics (II)}
\author{Natalia Gorobey, Alexander Lukyanenko}
\email{alex.lukyan@rambler.ru}
\affiliation{Department of Experimental Physics, St. Petersburg State Polytechnical
University, Polytekhnicheskaya 29, 195251, St. Petersburg, Russia}
\author{Inna Lukyanenko}
\email{inna.lukyanen@gmail.com}
\affiliation{Institut f\"{u}r Mathematik, TU Berlin, Strasse des 17 Juni 136, 10623
Berlin, Germany}

\begin{abstract}
Quantum Action Principle formulated earlier is used as a ground for a
probabilistic interpretation of one-particle relativistic quantum mechanics.
In this new approach the probability "flows" in the Minkowsky space being
dependent on an inner time parameter which we interpret as a particle life
time. The life time is determined as a function of observable parameters of
the real experiment by means of an additional condition of stationarity for
the quantum action.
\end{abstract}

\maketitle
\date{\today }

%\pacs{}

%\begin{multicols}{2}
%\narrowtext

%%%%%%%%%%%%%%%%%%%%%%%%%%%%%%%%%%%%%%%%%%%%%%%%%%%%%%%%%%%%%%%%%%%%%%%

%%\bigskip

\section{\textbf{INTRODUCTION}}

The subject of the present work is the problem of probabilistic
interpretation of relativistic quantum mechanics (RQM) (see, for example,
\cite{BD}). It arises first as the problem of probabilistic interpretation
of solutions of the simplest relativistic wave equation for one particle,
Klein-Gordon (KG) equation (velocity of light is equal unity) %%
\begin{equation}
\widehat{F}\psi \equiv \left( \theta _{\mu }\partial _{\mu }^{2}+\frac{m^{2}%
}{\hbar ^{2}}\right) \psi =0,  \label{1}
\end{equation}%
where $\theta _{\mu }=\left( +1,-1,-1,-1\right) $ is the signature of the
Minkowsky metrics, the summation is implied over repeated indices. But for
its solution one had to reject one-particle delivery of the problem. In
quantum field theory, which replaces RQM, the dynamical object is a quantum
field but not one particle. The reason to return to the problem in the
present work is a new formulation of quantum mechanics based on a quantum
action principle (QAP) \cite{GL}. In the work \cite{GL1} the attention was
turned to a possibility of probabilistic interpretation of one-particle RQM
in the framework of QAP. Instead of a wave function $\psi \left( x\right) $
considered as a solution of the Eq.(\ref{1}), the quantum dynamics in QAP is
described by a wave functional $\Psi \left[ x\left( c\right) \right] $ on
world lines $x_{\mu }\left( c\right) ,c\in \left[ 0,C\right] $ of a particle
in the Minkowsky space with fixed end points $x_{0\mu }\equiv x_{0\mu
}\left( 0\right) ,x_{1\mu }\equiv x_{1\mu }\left( C\right) $. The wave
functional has a natural probabilistic interpretation: $\left\vert \Psi %
\left[ x\left( c\right) \right] \right\vert ^{2}$ is a density of
probability of particle movement along a world line from a small
neighborhood of given world line $x_{\mu }\left( c\right) $. However, a
connection of the new framework with real measurements was not established.
To achieve this connection we must define the parameter $C$ as a function of
kinematical parameters of a real experiment. For this purpose, we have to
change the delivery of the scattering problem by introduction of wave
packets instead of plane waves for asymptotic states, and be more
considerable to experimental procedures of forming of initial state and
detection of particles. Notice that in non-relativistic quantum mechanics
the probabilistic interpretation of one-particle scattering problem needs
introduction of wave packets for asymptotic states \cite{FY}. In the new
approach to RQM the initial and final states of a particle are given by wave
packets in the Minkowsky space with corresponding parameters of space-time
coherence. The one-particle problem considered in the present work will be a
base of a more complicated problem of many--particle scattering.
Consideration of this problem in the framework of QAP will need a multy-time
description of dynamics \cite{GLL1},\cite{GLL2}.

\section{QUANTUM ACTION PRINCIPLE IN RELATIVISTIC MECHANICS}

We begin with a canonical form of the classical action of relativistic
particle, %%
\begin{equation}
I=\int\limits_{0}^{C}\left( p_{\mu }\overset{\cdot }{x}_{\mu }-\theta _{\mu
}p_{\mu }^{2}+m^{2}\right) dc,  \label{2}
\end{equation}%
where the high limit of integration $C$ has to be considered as an
independent dynamical parameter \cite{Fo}. It is defined by a condition of
stationarity for the action (2) calculated on solutions of classical
equations of motion and is proportional to the proper time of the particle
between given end points: %%
\begin{equation}
C=\frac{\sqrt{\theta _{\mu }\left( x_{1\mu }-x_{0\mu }\right) ^{2}}}{2m}.
\label{3}
\end{equation}%
In the new approach to RQM proposed in the present work the parameter $C$
will be defined after solution of all dynamical equations from a condition
of stationarity for a quantum action.

The difference of this new approach from the original one is, first of all,
in the operator realization of basic canonical variables. Now we introduce
their operator realization in a space of wave functionals as follows: %%
\begin{equation}
\widehat{x}_{\mu }\left( c\right) \Psi \equiv x_{\mu }\left( c\right) \Psi
,\,\,\,\,\,\,\, \widehat{p}_{\mu }\left( c\right) \Psi \equiv \frac{%
\widetilde{\hbar }}{i}\frac{\delta \Psi }{\delta x_{\mu }\left( c\right) },
\label{4}
\end{equation}%
where the variational derivative is defined by the equality %%
\begin{equation}
\delta \Psi =\int\limits_{0}^{C}\frac{\delta \Psi }{\delta x_{\mu }\left(
c\right) }\delta x_{\mu }\left( c\right) dc,  \label{5}
\end{equation}%
and the constant $\widetilde{\hbar }$ differs from the "ordinary" Plank
constant $\hbar $. For instance, its physical dimensionality is $Joule\times
\sec ^{2}/kg$ (in accordance with dimensionality of $C$). A connection
between two constants will be established latter. Let us introduce in the
space of wave functionals the Hermitian scalar product %%
\begin{equation}
\left( \Psi _{1},\Psi _{2}\right) \equiv \int \prod\limits_{c}d^{4}x\left(
c\right) \overline{\Psi }_{1}\left[ x\left( c\right) \right] \Psi _{2}\left[
x\left( c\right) \right] .  \label{6}
\end{equation}%
If the normalization condition $\left\vert \left\vert \Psi \right\vert
\right\vert \equiv \sqrt{\left( \Psi ,\Psi \right) }=1$ is fulfilled, then
the wave functional $\Psi \left[ x\left( c\right) \right] $ has a natural
probabilistic interpretation, namely, $\left\vert \Psi \left[ x\left(
c\right) \right] \right\vert ^{2}$ is a probability density of particle
movement along a world line from a small neighborhood of $x_{\mu }\left(
c\right) $. The operators defined by the Eq.(\ref{4}) are formally Hermitian
with respect to the scalar product (\ref{6}).

Let us turn to the formulation of QAP as a basic dynamical principle for
definition of a particle wave functional and corresponding quantum action.
Let us introduce an action operator by replacing in Eq. (\ref{2}) the basic
canonical variables by operators defined by the Eq.(\ref{4}): %%
\begin{equation}
\widehat{I}\equiv \int\limits_{0}^{C}\left[ \frac{\widetilde{\hbar }}{i}%
\overset{\cdot }{x}_{\mu }\left( c\right) \frac{\delta }{\delta x_{\mu
}\left( c\right) }+\widetilde{\hbar }^{2}\theta _{\mu }\frac{\delta ^{2}}{%
\delta x_{\mu }^{2}\left( c\right) }+m^{2}\right] dc.  \label{7}
\end{equation}%
This operator is formally Hermitian under condition that the product of
noncommuting operator multipliers in the first term under the integral is
symmetrizated. We formulate QAP as the eigenvalue problem for the action
operator, %%
\begin{equation}
\widehat{I}\Psi =\Lambda \Psi .  \label{8}
\end{equation}%
An eigenvalue $\Lambda $ and the corresponding eigenfunctional $\Psi \left[
x\left( c\right) \right] $ of the action operator depend on the end points $%
x_{1},x_{0}$, the invariant parameter $C$, and an infinite set of parameters
which define the state of motion of particle. The corresponding
eigenfunctional $\Psi \left[ x\left( c\right) \right] $ depends on the
complete set of these parameters too.

To look for a solution of the eigenvalue problem (\ref{8}), we introduce the
exponential representation of a wave functional %%
\begin{equation}
\Psi \left[ x\left( c\right) \right] =\exp \left\{ \frac{i}{\widetilde{\hbar
}}S\left[ x\left( c\right) \right] +R\left[ x\left( c\right) \right]
\right\} ,  \label{9}
\end{equation}%
and use a so called local approximation (\cite{GL}) for real functionals in
the exponent, %%
\begin{eqnarray}
S\left[ x\left( c\right) \right] &=&\int\limits_{0}^{C}s\left( c,x\left(
c\right) \right) dc,  \label{10} \\
R\left[ x\left( c\right) \right] &=&\int\limits_{0}^{C}r\left( c,x\left(
c\right) \right) dc.  \label{11}
\end{eqnarray}%
Notice that the local approximation is singular in QAP, in so far as the
repeated variational derivative in the action operator (\ref{7}) is
proportional to $\delta \left( 0\right) $. However, QAP in the local
approximation can be reduced to Schr\"{o}dinger wave equation \cite{GL} by
means of a proper regularization of (\ref{7}). It is achieved by division of
the interval $\left[ 0,C\right] $ into small parts of equal length $%
\varepsilon =C/N$, and approximation of a world line $x_{\mu }\left(
c\right) $ by a broken line with vertices $x_{n\mu }\equiv x_{\mu }\left(
n\varepsilon \right) $. Then the exponential functional (\ref{9}) can be
approximated by the product %%
\begin{equation}
\Psi \left[ x\left( c\right) \right] \simeq \prod\limits_{n=1}^{N-1}\psi
\left( n\varepsilon ,x_{n}\right) ,  \label{12}
\end{equation}%
where %%
\begin{eqnarray}
\psi \left( c,x\right) &=&\exp \chi \left( c,x\right) ,  \notag \\
\chi \left( c,x\right) &\equiv &\frac{i}{\hbar }s\left( c,x\left( c\right)
\right) +\varepsilon r\left( c,x\left( c\right) \right) ,  \label{13}
\end{eqnarray}%
and the relation %%
\begin{equation}
\widetilde{\hbar }=\varepsilon \hbar  \label{14}
\end{equation}%
was introduced (\cite{GL}). At the final stage, the limit $\varepsilon
\rightarrow 0$ is assumed and the product $\varepsilon r$ has to be
considered as a unit symbol. Formally, this limit is senseless, because $%
\widetilde{\hbar }\rightarrow 0$. It is a consequence of the accepted local
approximation. However, the new operator realization of basic canonical
variables (\ref{4}), as well as the action operator (\ref{7}), has definite
meaning in the discrete approximation. In the discrete approximation (\ref%
{12}), a wave functional is a function of many variables (the coordinates of
vertices $x_{n\mu }$) and the variational derivative can be approximated by
partial derivative as follows \cite{Fe}: %%
\begin{equation}
\frac{\delta \Psi }{\delta x_{\mu }\left( n\varepsilon \right) }\simeq \frac{%
1}{\varepsilon }\frac{\partial \Psi }{\partial x_{n\mu }}.  \label{15}
\end{equation}%
As a result, the action operator (\ref{7}) becomes a differential operator
with regularized second term under the integral. Considering the discrete
approximation as an intermediate stage, we obtain in the limit $\varepsilon
\rightarrow 0$ (\cite{GL}) %%
\begin{equation}
\frac{\widehat{I}\Psi }{\Psi }=\frac{\hbar }{i}\left. \chi \right\vert
_{0}^{C}+\int\limits_{0}^{C}\frac{Sch\psi \left( c,x\left( c\right) \right)
}{\psi \left( c,x\left( c\right) \right) }dc,  \label{16}
\end{equation}%
where %%
\begin{equation}
Sch\psi \left( c,x\left( c\right) \right) \equiv i\hbar \frac{\partial \psi
\left( c,x\left( c\right) \right) }{\partial c}-\widehat{F}\psi \left(
c,x\left( c\right) \right) .  \label{17}
\end{equation}%
The right hand side of the Eq.(\ref{16}) is equal to an eigenvalue of the
action operator if the expression (\ref{17}) equals zero for arbitrary world
line $x_{\mu }\left( c\right) $ with fixed end points. It is fulfilled for
any solution of Schr\"{o}dinger equation %%
\begin{equation}
i\hbar \frac{\partial \psi }{\partial c}=\widehat{F}\psi .  \label{18}
\end{equation}%
Incidentally, if the product in the first term under the integral in the Eq.(%
\ref{7}) is symmetrizated, then the eigenvalue is real and equals %%
\begin{equation}
\Lambda =\left. s\left( c,x\left( c\right) \right) \right\vert _{0}^{C},
\label{19}
\end{equation}%
where $s\left( c,x\left( c\right) \right) $ is the (real) phase function
defined by the Eq.(\ref{13}) for a solution $\psi \left( c,x\left( c\right)
\right) $. The latter will be called a wave function. Therefore, QAP in the
local approximation is reduced to the Schr\"{o}dinger equation (\ref{18})
with the time parameter $c\in \left[ 0,C\right] $, and the quantum action $%
\Lambda $ is defined by boundary values of real phase of the wave function.
The parameters, which define a quantum state of motion of particle,
mentioned above, are initial data of the Cauchy problem for Schr\"{o}dinger
equation (\ref{18}).

The notion of quantum action was introduced first by Dirac \cite{D} which
identified it with the (complex) phase of a wave function (\ref{13}) $\left(
\hbar /i\right) \chi $. We take only the real part of the phase. In fact,
now it is the ordinary definition of quantum action, but it was obtained
independently from QAP. Our formulation of QAP in terms of the eigenvalue
problem (\ref{8}) in a space of wave functionals differs from the Schwinger
quantum action principle \cite{Sch} which defines a quantum action as a
Hermitian operator in a space of wave functions. However, conditions of
stationarity of the quantum action with respect to inner parameters of the
system in both approaches are identical.

Schr\"{o}dinger equation (\ref{18}) can be obtained without QAP as a result
of formal application of the standard quantization procedure to the action (%
\ref{2}). Such quantum theory is formal because the parameter $c$ is not
measurable. In fact, this equation arises at an intermediate stage of
definition of the Feynman propagator (\cite{DeW},\cite{Go}), where at the
final stage integration of its solution $\psi \left( C,x\right) $ over the
length of a time interval $C$ with the measure $C^{-2}$ is assumed. However,
in the framework of QAP the integration is senseless, as far as the result
loses a direct connection with the original wave functional $\Psi \left[
x\left( c\right) \right] $, and, as a consequence, a possibility of
probabilistic interpretation given by QAP. Instead of this integration we
will fix the parameter $C$ by means of a quantum action stationarity
condition. Therefore, QAP in this case gives a connection of solutions of
Schr\"{o}dinger equation (\ref{18}) with real measurements.

Probabilistic interpretation of a wave functional $\Psi \left[ x\left(
c\right) \right] $ follows the probabilistic interpretation of the
corresponding wave function $\psi \left( c,x\right) $. Namely, $\left\vert
\psi \left( c,x\right) \right\vert ^{2}$ is a probability density of a
particle to be find at the moment of time $c$ in a small neighborhood of the
point $x$ in the Minkowsky space. This probabilistic measure has no direct
physical meaning until it is not connected with the real measurement. We
"bind" the parameter $c$ with measurements as follows: at the moment $c=0$ a
free particle springs up, and at the moment $c=C$ the particle disappears.
Before the moment $c=0$ the particle was in a bound state in a source (for
example, an electron in cathode), and after the moment $c=C$ it once again
comes in a bound state in a detector (Faradey cylinder). Therefore, $C$ is a
life time of a free particle. This interpretation is in accordance with the
Dirac interpretation of lover and upper limits of integrations in the action
(\ref{2}) \cite{D1}.

\section{PROBABILISTIC INTERPRETATION OF RELATIVISTIC QUANTUM MECHANICS}

Probabilistic interpretation of a wave function in ordinary quantum
mechanics is based on a differential conservation law of the probability,
which follows from Schr\"{o}dinger equation. In the new framework the
quantity $\left\vert \psi \left( c,x\right) \right\vert ^{2}$ obeys a
differential conservation law, as well, which is the consequence of the Eq.(%
\ref{18}). This probability "flows" in the Minkowsky space, but not in the
real one. If we shall fix the constant $C$ by a condition of stationarity of
the quantum action, the transition amplitude between initial and final
states of particle will become a function of observable quantities, and the
probability will be connected with real measurements. The quantum action
itself will be defined as a (real) phase of the transition amplitude.

In the preceding section we have defined the quantum action for arbitrary
solution of Schr\"{o}dinger equation (\ref{18}) as the difference (\ref{19})
of phases of the initial and final states. Here we correct this formal
definition, taking in place of a solution $\psi \left( C,x\right) $ of the
Cauchy problem a complex transition amplitude from a given initial state $%
\psi _{in}\left( x_{0}\right) $ in a source to a given final state $\psi
_{out}\left( x_{1}\right) $ in a detector. We consider the detection process
as a reduction of the final wave function $\psi \left( C,x\right) $ to a
state corresponding to the detector: %%
\begin{equation}
K\equiv \left\langle \psi _{out}\right. e^{-\frac{i}{\hbar }\widehat{F}%
C}\left. \psi _{in}\right\rangle .  \label{20}
\end{equation}%
The symmetry between the initial and final states presenting in the
amplitude (\ref{20}) will be need us latter for consideration of
annihilation processes in the framework of QAP.

Let us give the initial and final states of a particle. The particle springs
up in a source in a process (like an electron emission in a cathode), which
is localized in a finite domain of space-time. We correlate the initial
state with a wave packet in the Minkowsky space: %%
\begin{equation}
\psi _{in}\left( x_{0}\right) =A_{0}\exp \left[ -\frac{\left( x_{0\mu
}-X_{0\mu }\right) ^{2}}{2\sigma _{0\mu }^{2}}-\frac{i}{\hbar }\theta _{\mu
}p_{0\mu }x_{0\mu }\right] .  \label{21}
\end{equation}%
According to (\ref{21}), the source has the form of an ellipsoid in the
Minkowsky space (in a fixed reference frame) centered in the point $X_{0\mu }
$, and having space-time dimensions $\sigma _{0\mu }$. In the momentum
representation this state is described by a wave packet. It is centered in
the point $p_{0\mu }$ of a momentum space and has minimal dimensions $\hbar
/\sigma _{0\mu }$ admitted by Heizenberg uncertainly principle. We shall
call this state as the De-Broglie wave with the momentum $p_{0\mu }$ and the
parameters of coherence $\sigma _{0\mu }$. In accordance with the symmetry
demand, the final state has to be parameterized analogously. In this case
for the amplitude (\ref{20}) we obtain the following representation: %%
\begin{eqnarray}
K &=&A_{0}A_{1}\int d^{4}p\exp \left( -\frac{i}{\hbar }\theta _{\mu }p_{\mu
}^{2}C+\frac{i}{\hbar }m^{2}C\right) \times   \notag \\
&&\!\!\!\!\!\!\!\!\!\!\!\!\!\!\exp \left[ -\frac{\left( \sigma _{0\mu
}^{2}+\sigma _{1\mu }^{2}\right) }{2\hbar ^{2}}\left( p_{\mu }^{2}-2p_{\mu }%
\frac{\sigma _{0\mu }^{2}p_{0\mu }+\sigma _{1\mu }^{2}p_{1\mu }}{\sigma
_{0\mu }^{2}+\sigma _{1\mu }^{2}}\right) \right] \times   \notag \\
&&\!\!\!\!\!\!\!\!\!\!\!\!\!\!\exp \left( \frac{i}{\hbar }\theta _{\mu
}p_{\mu }\Delta X_{\mu }\right) ,  \label{22}
\end{eqnarray}%
where $\Delta X_{\mu }\equiv x_{1\mu }-x_{0\mu }$. The first exponent under
the integral is the evolution operator for the equation (\ref{18}) in the
momentum representation. However, in the present work we destroy the
symmetry between initial and final states, taking $\sigma _{1\mu }<<\sigma
_{0\mu }$. The process of detection of a particle (if we speak about a
particle) has not to be characterized by large space-time domain. The final
expression for the amplitude (\ref{20}) in this case is %%
\begin{eqnarray}
K &=&A\left( \prod\limits_{\mu }\sqrt{1+2i\hbar \frac{\theta _{\mu }C}{%
\sigma _{0\mu }^{2}}}\right) ^{-1}\times   \label{23} \\
&&\exp \left[ \frac{\sigma _{0\mu }^{2}}{2\hbar ^{2}}\frac{\left( p_{0\mu
}+i\hbar \theta _{\mu }\Delta X_{\mu }/\sigma _{0\mu }^{2}\right) ^{2}}{%
1+2i\hbar \theta _{\mu }C/\sigma _{0\mu }^{2}}+\frac{i}{\hbar }m^{2}C\right]
.  \notag
\end{eqnarray}%
All multipliers which do not depend on the parameters $\Delta X_{\mu }$, and
$C$ are included into a common constant $A$. Summation over the indixe $\mu $
in the exponent is assumed as usually.

Therefore, in our experiment the particle-wave duality is transparent: a
De-Brogile wave is prepared in a source, and a particle is registered in a
detector. The probability of the process equals $\left\vert K\right\vert
^{2} $, under condition that the life time of particle $C$ \ is determined.
The phase of the transition amplitude $K$ plays the role of the quantum
action. From (\ref{23}) one obtains that %%
\begin{eqnarray}
\Lambda &=&m^{2}C-\left[ \frac{\hbar }{2}\arctan \frac{2\hbar \theta _{\mu }C%
}{\sigma _{0\mu }^{2}}+\right.  \label{24} \\
&&\left. \theta _{\mu }\frac{\left( p_{0\mu }^{2}-\hbar ^{2}\Delta X_{\mu
}/\sigma _{0\mu }^{4}\right) C-p_{0\mu }\Delta X_{\mu }}{1+4\hbar
^{2}C^{2}/\sigma _{0\mu }^{4}}\right] .  \notag
\end{eqnarray}%
The classical limit $\hbar =0$ of the quantum action (\ref{24}) coincides
with the original classical action (\ref{2}) for a particle moving with the
constant momentum $p_{0\mu }$ between two given end points: %%
\begin{equation}
\Lambda _{0}=\theta _{\mu }p_{0\mu }\Delta X_{\mu }-\left( \theta _{\mu
}p_{0\mu }^{2}-m^{2}\right) C.  \label{25}
\end{equation}%
The condition of stationarity of the classical action (\ref{25}) with
respect to $C$ gives the classical constraint for the initial momentum of
particle, %%
\begin{equation}
\theta _{\mu }p_{0\mu }^{2}-m^{2}=0,  \label{26}
\end{equation}%
which was not supposed from the beginning. In classical relativistic
mechanics the Eq.(\ref{26}) determines $C$, if we take into account the
classical relation: %%
\begin{equation}
p_{0\mu }=\frac{\Delta X_{\mu }}{2C}.  \label{27}
\end{equation}%
In quantum theory this relation in a precise form is absent but it can be
considered as an experimental result. Indeed, the real part of the
transition amplitude (\ref{24}) which defines the probability "flow" is
proportional to the exponent %%
\begin{equation}
\exp \left[ -\frac{\left( \Delta X_{\mu }-2p_{0\mu }C\right) ^{2}}{2\left(
\sigma _{0\mu }^{2}+4\hbar ^{2}C^{2}/\sigma _{0\mu }^{2}\right) }\right] ,
\label{28}
\end{equation}%
which predicts maximal probability of the registration of a particle for
detectors localized in accordance with (\ref{27}). Therefore, one can expect
that in the classical limit the sought-for stationary value of $C$ will be
equal (\ref{3}).

However, the complete quantum action (\ref{24}) gives us a possibility to
fix $C$ definitely by means of the stationarity condition, %%
\begin{equation}
\frac{\partial \Lambda }{\partial C}=0,  \label{29}
\end{equation}%
without any additional equations of motion. Accounting quantum corrections
to the classical action, %%
\begin{equation}
\Lambda =\Lambda _{0}+\hbar ^{2}\Lambda _{2}+...,  \label{30}
\end{equation}%
where %%
\begin{eqnarray}
\Lambda _{2} &=&-\left[ 4\theta _{\mu }\left( p_{0\mu }\Delta X\mu -p_{0\mu
}^{2}C\right) \frac{C^{2}}{\sigma _{0\mu }^{4}}+\right.  \notag \\
&&\left. \theta _{\mu }\left( \frac{1}{\sigma _{0\mu }^{2}}-\frac{\Delta
X_{\mu }^{2}}{\sigma _{0\mu }^{4}}\right) C\right] ,  \label{31}
\end{eqnarray}%
one can obtain a quasi-classical decomposition of $C$. In the zero
approximation $\left( \hbar =0\right) $ the stationary value of $\ C$ equals
\begin{equation}
C=\frac{8\sum\limits_{\mu }\theta _{\mu }p_{0\mu }\Delta X_{\mu }/\sigma
_{0\mu }^{4}\pm \sqrt{D}}{24\sum\limits_{\mu }\theta _{\mu }p_{0\mu
}^{2}/\sigma _{0\mu }^{4}},  \label{32}
\end{equation}%
where %%
\begin{eqnarray}
D &\equiv &64\left( \sum\limits_{\mu }\theta _{\mu }p_{0\mu }\Delta X_{\mu
}/\sigma _{0\mu }^{4}\right) ^{2}+  \label{33} \\
&&48\left( \sum\limits_{\mu }\theta _{\mu }p_{0\mu }^{2}/\sigma _{0\mu
}^{4}\right) \left( \sum\limits_{\mu }\theta _{\mu }\left( \frac{1}{\sigma
_{0\mu }^{2}}-\frac{\Delta X_{\mu }^{2}}{\sigma _{0\mu }^{4}}\right) \right)
.  \notag
\end{eqnarray}%
Here we first write the summation symbol. To select the sign in (\ref{32})
one can use the classical limit (\ref{3}). If detectors are sufficiently
distant from the source, so that %%
\begin{equation}
\sum\limits_{\mu }\theta _{\mu }\frac{\Delta X_{\mu }^{2}}{\sigma _{0\mu
}^{4}}>>\left\vert \sum\limits_{\mu }\theta _{\mu }\frac{1}{\sigma _{0\mu
}^{2}}\right\vert ,  \label{34}
\end{equation}%
and the classical relation (\ref{27}) is fulfilled (that is an experimental
result), then the choice of the sign $+$ entails the correct classical limit
(\ref{3}). Quantum corrections to $C$ we obtain by accounting higher order
terms in the decomposition (\ref{30}). Substituting the stationary value $C$
in the transition amplitude (\ref{23}), we obtain a function only of
kinematical parameters of the experiment: $K\left( p_{0},\Delta X\right) .$
It is this amplitude that will be a ground of our probabilistic
interpretation of RQM.

Consider the nonrelativistic limit of the amplitude $K\left( p_{0},\Delta
X\right) $. Taking $p_{00}\simeq m>>\left\vert p_{0k}\right\vert $, and $%
\Delta X_{0}>>\left\vert \Delta X_{k}\right\vert $, we obtain from (\ref{32}%
) %%
\begin{equation*}
C\simeq \frac{\Delta X_{0}}{2m},
\end{equation*}%
i.e. in the non-relativistic limit the life time of particle $C$ is
proportional to the ordinary Newtonian time. Then the part of the exponent (%
\ref{28}), corresponding to $\mu =0$, may be omitted, and the remaining part
of the exponent describes a packet moving with the Newtonian time $t\equiv
\Delta X_{0}$ in the ordinary three-dimensional space. According to (\ref{25}%
), the phase of the packet in that limit is $mt$, so that the packet obeys
Schr\"{o}dinger equation with the Hamiltonian %%
\begin{equation*}
-\left( m+\frac{\hbar ^{2}}{2m}\Delta \right) ,
\end{equation*}%
as it must be in the non-relativistic limit.

Coming back to the main problem, let as finish description of the
experiment. As it is accepted in low energy electron diffraction, let a
source of particles is placed in the interior of a large screen $\Theta $,
so that the relation (\ref{34}) is fulfilled. Therefore, the domain of the
Minkowsky space, where a particle springs up, is placed into a space-time
cilinder, which space section coincides with the interior of the screen, and
having infinite length in the time direction. If whole inner surface of the
screen is filled by detectors, a particle springing up in the source, will
be detected sooner or later by someone of detectors. A ground for our
assurance is the conservation law for of probability "flow" mentioned above.
Indeed, the center of a wave packet, which springs up in the source at the
moment $c=0$, in accordance with (\ref{28}), moves along a straight line in
the ordinary space: %%
\begin{equation*}
X_{k}\left( c\right) =X_{0k}+\frac{p_{0k}}{m}c,
\end{equation*}%
the packet itself expands. Therefore, the packet "leaves" the cylinder with
increasing the parameter $c$. It may be expressed in another words: the
probability of a world line beginning in the source to remain in the
interior of the cylinder, has zero limit when $c\rightarrow \infty $. Now,
let the transition amplitude $K\left( p_{0},\Delta X\right) $ obeys the
normalization condition %%
\begin{equation}
\int \limits_{\Theta }d\Omega \int \limits_{0}^{\infty }d\left( \Delta
X_{0}\right) \left\vert K\right\vert ^{2}=1,  \label{35}
\end{equation}%
where the first integral is taken over angular variables on the screen $%
\Theta $ surface, and the second one - over the whole waiting time of
snapping into action someone of detectors. Let us stress that time is an
equal (along with the space coordinates) parameter of that distribution of
probability. Finiteness of the integral over the time variable $\Delta X_{0}$
is ensured by a multiplier in front of the exponent in (\ref{23}), which
decreases %%diminishes
like $\Delta X_{0}^{-2}$ when $\Delta X_{0}\rightarrow \infty $, so that the
normalization condition (\ref{35}) is meaningfull. After that, the quantity $%
\left\vert K\left( p_{0},\Delta X\right) \right\vert ^{2}$ will play the
role of a probability density of snapping into action someone of detectors
in the screen surface (some time or other).

We have considered in the present work the simplest experiment with a free
relativistic particle. In the presence of an external electromagnetic field
the action (\ref{2}) has to be replaced by the following one: %%
\begin{equation}
I=\int\limits_{0}^{C}\left[ p_{\mu }\overset{\cdot }{x}_{\mu }-\theta _{\mu
}\left( p_{\mu }-eA_{\mu }\right) ^{2}+m^{2}\right] dc.  \label{36}
\end{equation}%
Formulation of QAP and corresponding scattering problem for the action (\ref%
{36}) has no principal difficulties, it may be realized in the framework of
perturbation theory on the electric charge $e$. Principal for the new
approach is the problem of interaction of a particle with quantum
electromagnetic field. It will be considered in the framework of QAP in a
subsequent work.

\section{\textbf{CONCLUSIONS }}

Therefore, quantum action principle gives a possibility of proper
probabilistic interpretation of relativistic quantum mechanics in the simple
one-particle scattering experiment. We consider the present work as a ground
for consideration of multi-particle scattering problem in the framework of
QAP.

We are thanks V. A. Franke and A. V. Goltsev for useful discussions.

%%%\noindent $^{\ast }$ E-mail address: alex.lukyan@rambler.ru

%%%\noindent $^{+}$ E-mail address: inna.lukyan@mail.ru

%\begin{references}

\end{document}